\documentclass[5p]{elsarticle}

\usepackage{lineno,hyperref}
\modulolinenumbers[5]

\journal{Journal of \LaTeX\ Templates}

\usepackage{graphicx}   %       for graphics
\usepackage{latexsym}   %       for special symbols
\usepackage{hyperref}   %       for hyperlinks

\begin{document}
%==============================================================================

\begin{frontmatter}
\title{Conserved charge fluctuations are not conserved during the hadronic phase}

\author{J. Steinheimer$^1$, V. Vovchenko$^{1,2,3}$, J. Aichelin$^{1,4}$, M. Bleicher$^{1,2}$, and H. St\"ocker$^{1,2,5}$}

\address{$^1$ Frankfurt Institute for Advanced Studies, Ruth-Moufang-Str. 1, 60438 Frankfurt am Main, Germany}
\address{$^2$ Institut f\"ur Theoretische Physik, Goethe Universit\"at Frankfurt, Max-von-Laue-Strasse 1, D-60438 Frankfurt am Main, Germany}
\address{$^3$ Department of Physics, Taras Shevchenko National University of Kiev, 03022 Kiev, Ukraine}
\address{$^4$ SUBATECH, UMR 6457, Universit\'{e} de Nantes, Ecole des Mines de Nantes, IN2P3/CNRS. 4 rue Alfred Kastler, 44307 Nantes cedex 3, France}
\address{$^5$ GSI Helmholtzzentrum f\"ur Schwerionenforschung GmbH, D-64291 Darmstadt, Germany}
%\date{February 28, 2014}

\begin{abstract}
We study the correlation between the distributions of the net-charge, net-kaon, net-baryon and net-proton number at hadronization and after the final hadronic decoupling by simulating ultra relativistic heavy ion collisions with the hybrid version of the ultrarelativistic quantum molecular dynamics (UrQMD) model. We find that due to the hadronic rescattering these distributions are not strongly correlated.
The calculated change of the correlation, during the hadronic expansion stage, does not support the recent paradigm, namely that the measured final moments of the experimentally observed distributions do give directly the values of those distributions at earlier times, when the system had been closer to the QCD crossover.
\end{abstract}

%\pacs{25.75.-q, 25.75.Gz, 24.10.Lx}

%\maketitle

\end{frontmatter}

In this paper we will question whether the fluctuations observed in high energy nuclear collisions can be related to the fluctuations 
close to the QCD pseudo critical line from lattice gauge calculations, an idea which has been recently advanced by several authors \cite{Ejiri:2005wq,Friman:2011pf,BraunMunzinger:2011dn,Karsch:2010ck}.\\
The collision of heavy ions at relativistic energies allows for the creation of ultra dense baryonic matter at very high energy densities and temperatures. It is expected from quantum chromo dynamics (QCD) that matter undergoes a transition to a novel state called the quark-gluon-plasma (QGP) \cite{Stephanov:1998dy} at high temperatures and even into a quarkyonic \cite{McLerran:2007qj} or a color-flavor-locked phase \cite{Alford:2007xm} at high baryon densities and moderate temperatures. 
Event-by-event fluctuations of conserved charges, measured in relativistic nuclear collisions, moved recently in the center of attention in the search of a possible first order phase transition and critical endpoint in the phase diagram of QCD \cite{Stephanov:2008qz,Koch:2008ia}.
Higher order cumulants of the net-charge distributions (related to the susceptibilities) are sensitive probes of the underlying equation of state and hence of the phase structure.
However, they are also very sensitive to other effects which are not related to the underlying physics of interest. Such effects include the correct understanding of efficiency and acceptance effects in experimental setups \cite{Bzdak:2012ab,Bzdak:2016qdc,Kitazawa:2016awu}, influences from cluster formation \cite{Feckova:2015qza}, influence of conservation laws \cite{Begun:2004gs,Bzdak:2012an}, corrections due to the finite size of the system created in relativistic nuclear collisions \cite{Gorenstein:2008et}, fluctuations of the system volume~\cite{Gorenstein:2011vq,Sangaline:2015bma} and fluctuations present in the initial state of the collision \cite{Spieles:1996is}.
Experiments have reported on measurements of conserved charge fluctuations (see e.g. \cite{Tarnowsky:2012vu,Xu:2014jsa,Adamczyk:2014fia,Adamczyk:2013dal}), but up to date most studies compare the cumulants computed on the lattice or within grand canonical effective models such as the hadron resonance gas (HRG) model, directly to the experimental data~\cite{Borsanyi:2014ewa,Ding:2015ona,Alba:2014eba}. 
It is important to note that in fact experiments can measure directly only the net-electric charge distribution and, thus, one relies on measurements of the net-proton and net-kaon distributions as proxies for the net-baryon and net-strangeness distributions \cite{Noronha-Hostler:2016rpd}. As was shown by Shuryak and Stephanov, by solving the Fokker-Planck equation for hadronic rescattering, the hadronic phase does change the long range charge correlations \cite{Shuryak:2000pd}. Several other studies have also found that in fact the elastic and inelastic scattering in the hadronic phase, which follows the hadronization process in nuclear collisions, may have a significant impact on the observed cumulants \cite{Kitazawa:2012at,Asakawa:2000wh,Jeon:2000wg}, and may further blur the correlation of particle number fluctuations and net-charge fluctuations. Quantitative statements are, however, difficult as they require a microscopic transport treatment of the hadronic rescattering phase (see e.g. \cite{Konchakovski:2005hq}).

We will show in this letter that hadronic rescattering, based on known reaction cross sections, decorrelates strongly the distributions of conserved charges between hadronization and kinetic freeze-out. Thus the experimentally observed moments of this distribution are only weakly correlated with the moments of the distribution at hadronization, i.e. at the boundary between the QGP and the hadronic world.

\section{The model}

For our investigations we will utilize the newest version of the UrQMD hybrid model \cite{Petersen:2008dd}, which employs the UrQMD hadronic transport model to simulate the non-equilibrium initial and final phases of the nuclear reactions. In the UrQMD model, all charges (baryon number, strangeness and electric charge) are exactly conserved globally and for every individual scattering which is an important feature for any realistic model used to study correlations and fluctuations. The bulk evolution of the hot and dense phase is simulated by a 3+1 dimensional ideal fluid dynamics code. The fluid dynamical equations for conservation of energy-momentum as well as the conservation of the net-baryon number current are solved by the SHASTA algorithm \cite{Rischke:1995ir}.
The transition from the initial non-equilibrium phase to the fluid dynamical evolution is done at a time $t_{\rm INI}$ defined as the interpenetration time, the earliest instant at which local equilibration can be achieved (with the constraint that $t_{\rm INI}\ge 0.5$ fm/c).
\begin{equation}
t_{\rm INI}=\frac{2 R}{\gamma_{\rm CM} \  v}=\frac{2 R}{\sqrt{\gamma_{\rm CM}^2-1}}
\end{equation}
where R is the radius of the nucleus and $\gamma_{CM}$ is the Lorentz gamma factor of the two nuclei in their center of mass frame.
The transition back to discrete particles from the fluid dynamical fields is done in the standard fashion, by employing the Cooper-Frye prescription \cite{Cooper:1974mv} on a pre-defined hypersurface. In this work we will employ the iso-energy density hypersurface \cite{Huovinen:2012is} of $e_{CF}\approx$~350 MeV/fm$^3$. To ensure global energy conservation as well as conservation of the different net-charges (electric, baryon number and strangeness) we also use a rejection method during the Monte-Carlo sampling of the Cooper-Frye equation (eq. (\ref{cooper_frye})).
Hadrons which have been produced at the hypersurface 
interact during the hadronic rescattering phase in the UrQMD model \cite{Bass:1998ca,Bleicher:1999xi}.
To describe this stage we use the cascade version of the code, where all interactions between hadrons are determined by a geometric criterion 
tied to their specific scattering cross section. The model includes close to 60 different baryonic species and their anti-particles as well as about 40 mesonic states.

\begin{figure}[t]	%       -----------------------------------------
\includegraphics[width=0.5\textwidth]{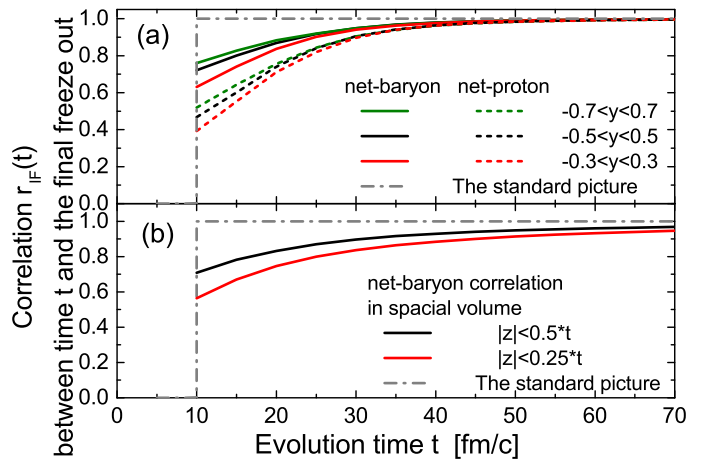}
\caption{[Color online] (a) Correlation of the finally observed net-proton and net-baryon number in different rapidity intervals with an earlier time $t$ in that same rapidity interval.
Results for central collisions of Au+Au at a beam energy of $\sqrt{s_{\rm NN}}= 200$~GeV are shown. (b) Same is in (a), just in this case we show the correlation for a defined spatial volume, constrained by $z<v_z \cdot t$ rather than rapidity.
}\label{f2}
\end{figure}		%       ----------------------------------------- 

The corresponding momentum dependent scattering cross sections are taken from experiment where known and from the additive quark model otherwise. 
The reactions implemented in the model include elastic $2\leftrightarrow 2$ scatterings as well as inelastic reactions like resonance excitations and decays, strangeness exchange and annihilation reactions.
A detailed analysis of the reactions in the final hadronic stage of an ultra relativistic heavy ion collision can be found in \cite{Steinheimer:2017vju}. We like to note that this kind of hybrid model, employing UrQMD as afterburner, has been used to successfully describe a wealth of experimental results including particle multiplicities, spectra, flow and HBT observables (see e.g. \cite{Werner:2010aa,Song:2013qma,Knospe:2015nva,Ryu:2012at,Steinheimer:2012rd}).

\begin{figure}[t]	%       -----------------------------------------
\includegraphics[width=0.5\textwidth]{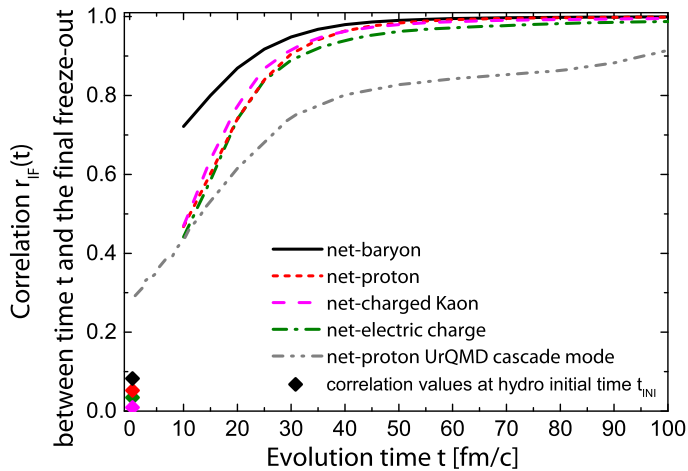}
\caption{[Color online] Correlation of the net-charge, net-kaon, net-proton and net-baryon number in the rapidity interval $-0.5<y<0.5$ with an earlier time $t$ in that same rapidity interval.
}\label{f1}
\end{figure}		%       ----------------------------------------- 

As we intend to study the effects of the hadronic rescattering on fluctuation measures as the cumulants of the different net-charge distributions,
we first have to discuss some specific difficulties associated with the current state-of-the-art fluid dynamical models. 
In the usual framework of fluid dynamical models, as well as of the hybrid models such as the one employed in this study, the transition
from the fluid dynamical fields to a finite and discrete number of particles is done by a Monte-Carlo sampling of the Cooper-Frye equation on a specific hypersurface $\sigma_\mu$
\begin{equation}
\label{cooper_frye}
E \frac{dN}{d^3p}=\int_\sigma f(x,p) p^\mu d\sigma_\mu \, .
\end{equation}

It is important to note that the single particle distribution functions $f(x,p)$ which enter the Cooper-Frye formula are usually assumed to be the grand canonical Fermi or Bose distributions for a hadron resonance gas at a given temperature $T$ and chemical potential $\mu_i$. 
These single-particle distribution functions are generally uncorrelated, as expected for a non-interacting HRG, and, since the particle number in a given local computational cell is much smaller than 1, the
particle number distributions locally follow the Poisson distribution \footnote{If global conservation laws are enforced the distributions are actually multinomial or binomial}. 
It is therefore obvious that, using this standard method of particlization, one cannot obtain a net-charge distribution which is significantly different from a Poisson distribution except for effects stemming from the global conservation of net-charges, as enforced by the particle production scheme employed in the UrQMD-hybrid model, as well as long range correlations from inhomogeneities in the density distribution if a phase transition has occurred \cite{Steinheimer:2012gc} (which is presently not included in the fluid dynamical model). 
Thus, the comparison of the (normalized) cumulants 
before and after the hadronic rescattering phase
will not lead to new insights as in this case one starts with a Poisson/Binomial distribution. 

In the current paper we will take a different approach to make qualitative and quantitative statements. 
We will set-up a time-dependent correlation function to explore the diffusion of conserved charges during the hadronic phase.
Following the standard definition of the correlation coefficient this function reads as
\begin{equation}
	r_{\rm IF}(t)= \frac{\sum\limits_{n} (I_n(t) - \overline{I}(t)) (F_n - \overline{F})}{\sqrt{\sum\limits_{n} (I_n(t) - \overline{I}(t))^2 \sum\limits_{n} (F_n - \overline{F})^2}},
\end{equation}
where $I_n(t)$ denotes the number of a given charge (or a net-particle number) in a given rapidity and momentum window at the time $t$ in the event number $n$. $F_n$ is the final number of that charge after all interactions have ceased, in the same event. Finally, the $\overline{I}(t)$ and $\overline{F}$ are the corresponding averages. The sum runs over all events in the sample. We want to make clear that this correlation relates the net charge at a given fixed time $t$ with the final net-charge observed in the same acceptance. Since we use an iso-energy density hypersurface the definition of a fixed time correlation does not make sense at a time which does not include all particles emitted through the hypersurface. Therefore we will mainly discuss this correlation for times $t$ that are larger than the latest emission from the hypersurface that enters in the Cooper-Frye sampling. In particular that means we mainly discuss times later than $t=10$ fm/c at which point essentially all hadrons have been produced.

It is evident that if the net-charge in a given bin at a given time $t$ is perfectly correlated with the net-charge at the end of the evolution, then the value of this correlation function $r_{\rm IF}(t)$ will be equal to 1 whereas if the net-charge number at given time $t$ is completely uncorrelated with the final value, then the $r_{\rm IF}(t)$ will be equal to zero.

\begin{figure}[t]	%       -----------------------------------------
\includegraphics[width=0.5\textwidth]{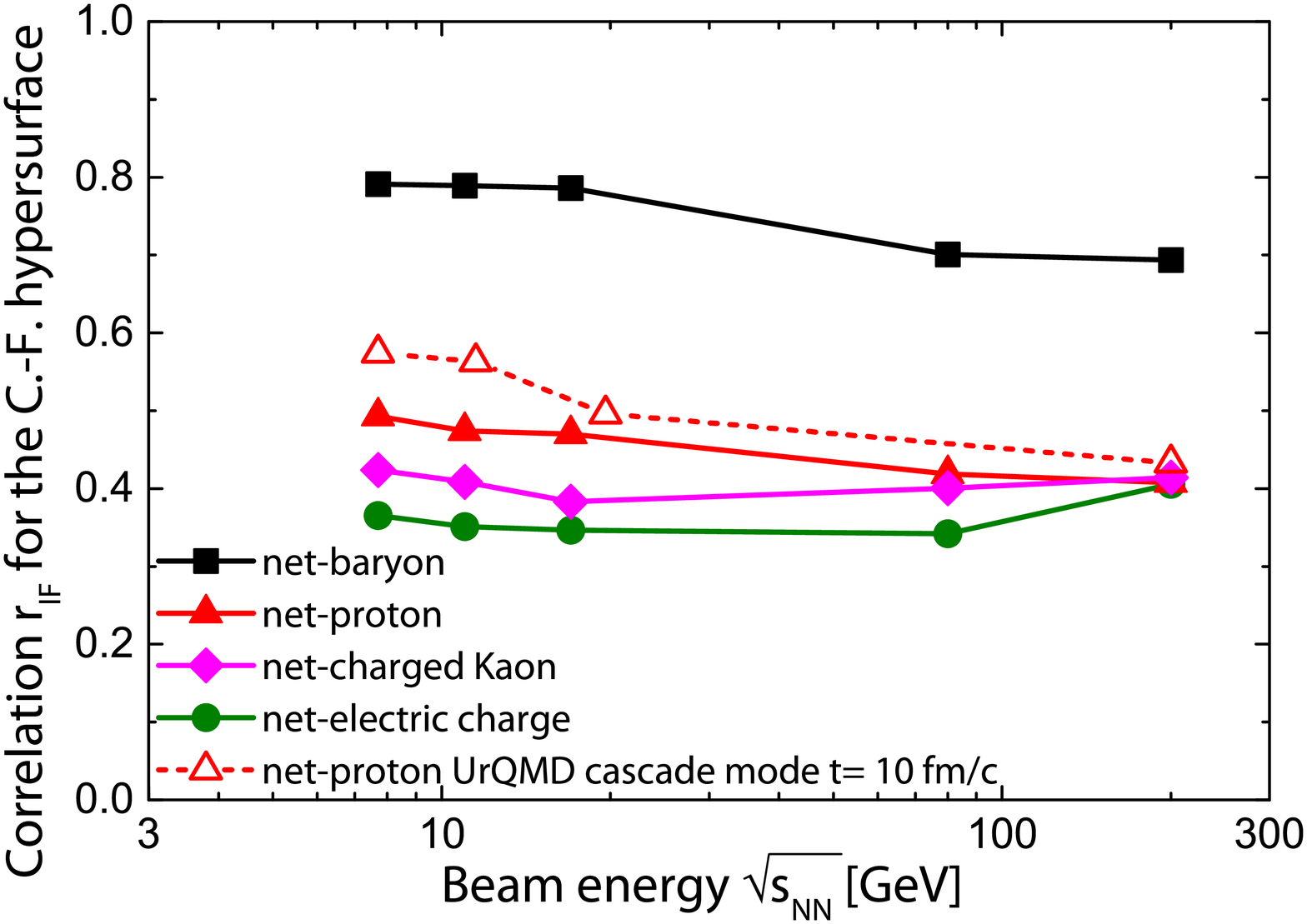}
\caption{[Color online] Correlation of the finally observed net-charge, net-kaon, net-proton and net-baryon number in the rapidity interval $-0.5<y<0.5$, at the Cooper-Frye hypersurface, with the final net-charge density as a function of beam energy.}\label{f3}
\end{figure}		%       ----------------------------------------- 

In the standard scenario \cite{Karsch:2010ck}, which is employed in most comparisons of the (normalized) cumulants with the HRG or lattice QCD results one essentially assumes that this correlation is equal to zero before the chemical freeze-out of hadrons and then instantly jumps to one at the chemical freeze-out, a value which does not change until the cumulants are observed by experiment.
This scenario is shown in figure \ref{f2} by a gray dash dotted line for an exemplary freeze-out time of 10 fm/c.
In such a scenario the final value of the cumulants correspond to that at chemical freeze-out, without any information about the QGP phase before freeze-out or about the hadronic rescattering which follows hadronization. 

\section{Results}

We start by studying the correlation of different net-charges in central ($b<3.4$ fm) collisions of gold nuclei at beam energy of $\sqrt{s_{\rm NN}}= 200$ GeV, as measured at the Relativistic Heavy Ion Collider (RHIC) at the Brookhaven National Laboratory (BNL). We will investigate the correlation of the net-electric charge, the net-kaon number, the net-baryon number and the net proton number within our model. For the net-proton and net-baryon number we also apply a cut in transverse momentum of $0.3 < p_{T} < 2.0$ ~GeV, which corresponds approximately to the relevant experimental acceptance window. All these quantities, except the net-baryon number which is inaccessible with the current detectors, have been studied in experiments at the RHIC.

\begin{figure}[t]	%       -----------------------------------------
\includegraphics[width=0.5\textwidth]{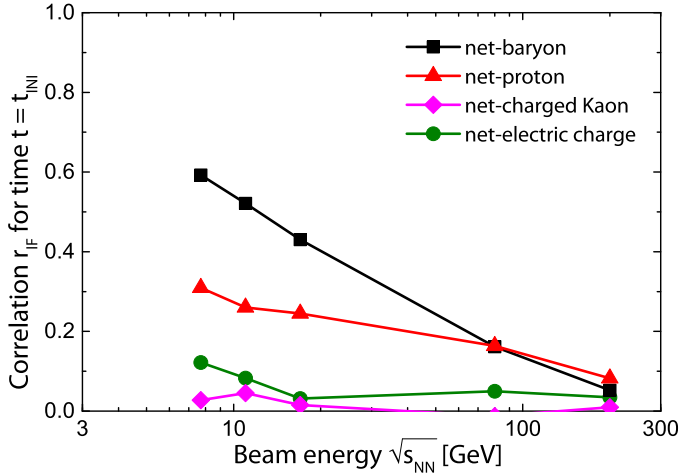}
\caption{[Color online] Correlation of the net-charge, net-kaon, net-proton and net-baryon number in the rapidity intervals $-0.5<y<0.5$ at the initial hydro time with the final net-charge density in that same rapidity interval, as a function of beam energy. Note that the addition decorrelation, with respect to figure \ref{f3} stems from the enforced local equilibration in the fluid dynamical evolution and the Cooper-Frye sampling at its end.}\label{f4}
\end{figure}		%       ----------------------------------------- 

To estimate the effect of the full local equilibration on the decorrelation of the different charges we will also contrast the hybrid model simulations with results obtained from the standard UrQMD model in its cascade version. In the cascade version of the model, the system remains essentially out of local equilibrium for a long time and, due to the finite cross sections, will retain more information and therefore show a stronger correlation with the early times.

First we focus on the influence of the size of the rapidity acceptance window, an effect which has also been discussed in a simplified picture \cite{Ohnishi:2016bdf}. 
Figure \ref{f2} (a) depicts the correlation of the net-baryon number and net-proton number as a function of time for central Au+Au reactions at $\sqrt{s_{\rm NN}}= 200$ GeV. We present the correlation for times $t> 10$ fm/c. The reason we choose to show the correlation only after 10 fm/c is that at this point
all hadrons have been emitted from the fluid dynamical phase. At an earlier time part of the system would
still be in the fluid dynamical phase and the calculation is indeed less meaningful. Furthermore the net-proton number cannot be defined during the deconfined phase which dominates the dynamics at earlier times and during the time period of hadron emission through the Cooper-Frye hypersurface. 

A dependence on the size of the rapidity window is clearly present. As expected, a larger acceptance window leads to a stronger correlation of the initial and final net-charge number in a single event. We also observe a stronger correlation of the net-baryon number with the finally observed number than for the net-proton number, which is also expected, as protons can easily exchange iso-spin and become neutrons during the rescattering phase, e.g. through the excitation and decay of a $\Delta$ resonance. To make sure that the de-correlation in coordinate space is of similar magnitude as the de-correlation in momentum space we also show $r_{\rm IF}(t)$ for net-baryons in a given spatial volume which is defined by the longitudinal boundary $|z|< v_z \cdot t$, where $v_z$ is the maximal velocity a particle can have and still belong to the rapidity interval of $|y|<0.5$ or $|y|<0.3$ respectively.

Figure \ref{f1} shows the dependence of the correlation $r_{\rm IF}(t)$, for different net-charge and net-particle numbers, as a function of time. 
Unlike in the standard scenario assuming instant freeze-out, the correlation function strongly depends on the duration of the rescattering phase for all considered quantities. At very late times ($t>40$ fm/c) the correlation function approaches 1, which is expected as the system has essentially frozen out and interactions have ceased by that time. However, the correlation changes rapidly in the time interval between particle production (on the Cooper-Frye hypersurface) at around 10 fm/c and decoupling of the system at about 25 fm/c. At the time of interest for most studies, namely the particle production time, the correlation is approximately 0.5 for most net-charges. Only the net-baryon number correlation is slightly stronger, as discussed above. However, the quantity that is measured experimentally, namely the net-proton number, shows a similar correlation as the net-charge, as significant iso-spin changing reactions take place during the rescattering phase. 
One should keep in mind that the fact that the correlation still changes at rather late times does not mean that the lifetime of the hadronic phase is on the order of 40 fm/c. At a time of 20fm/c about $90\%$ of all 2-body reactions have already occurred. After that we observe mainly resonance decays. These decays of course also de-correlate the number of charges in a given acceptance window. Therefore the slow convergence of the correlation is due to the long lifetime of the resonances and not to some late rescattering.
The net protons in the case of the full cascade UrQMD simulation show a larger initial correlation and an even slower increase of the correlation function compared to the hybrid simulation, hinting to a smaller number of scatterings in the early high density phase and an extended emission period. This is likely due to the smaller radial flow in the cascade simulation due to the lack of scatterings in the early stage of the collision.

There are two crucial points in time at which we want to study the correlation $r_{\rm IF}(t)$ in more detail. One is the time at which hadrons are produced according to chemical equilibrium distributions. At this point, at the transition between the hadronic and quark phase, the hadrons are sampled from the transition (hadronization) hypersurface according to the Cooper-Frye equation. In fact since the iso-energy hypesurface does not correspond to a fixed time we would not compare the net charge at two different fixed times, but on the hypersurface volume and in the final state. The second one is an even earlier time at which the fluid dynamical evolution starts. The second point is of interest because one can identify possible correlations which survive the fluid dynamical evolution. This can be, for example, correlations which stem from the initial stopping of the participant nucleons and which may have an influence on the final net-charge distributions.
Therefore we show in figure \ref{f3} the correlation $r_{\rm IF}(t=t_{\rm CF})$ which is defined as the correlation between the net-particle number going through the Cooper-Frye hypersurface and the final state particle number. Again this correlation does not compare two fixed times but two 4-volumes in which the relevant quantities are conserved globally. The correlation between the Cooper-Frye surface and the final state apparently shows only a very weak beam energy dependence down to $\sqrt{s_{\rm NN}}= 7$ GeV. This means that the importance of the hadronic rescattering does not change much within the beam energy range of the beam energy scan program at RHIC.
The results from the standard cascade UrQMD simulations show a similar correlation for all considered beam energies at $t=10$~fm/c. More important the beam energy dependence of the correlation $r_{\rm IF}(t=t_{\rm INI})$, 
defined as the correlation between the net-particle number at the starting time of the fluid dynamical simulation and the final particle number, shows a clear energy dependence in figure \ref{f4}. The results shown in figure \ref{f4} are interesting as they make clear that even if the system goes through a fluid dynamic phase, where one assumes local equilibrium, some information of the initial state does survive the whole dynamics. In a naive scenario where one assumes particle production from a thermal heat bath, no information of the initial state on the final net-charge distribution should survive. We show that in fact some information survives due to the fact that charges are globally conserved and the fluid dynamical evolution also conserves the local energy-momentum and baryon number current.
The correlation increases with decreasing beam energy. While at the top RHIC energy, essentially no information on baryon stopping survives the fluid dynamical phase, at low energies the conservation of the stopped baryon number throughout the fluid dynamical evolution generates a significant correlation. In general one would expect that the initial mapping of the conserved charges on the fluid dynamical grid would de-correlate the charges and the subsequent expansion would do so further and our findings confirm this expectation. 

\begin{figure}[t]	%       -----------------------------------------
\includegraphics[width=0.5\textwidth]{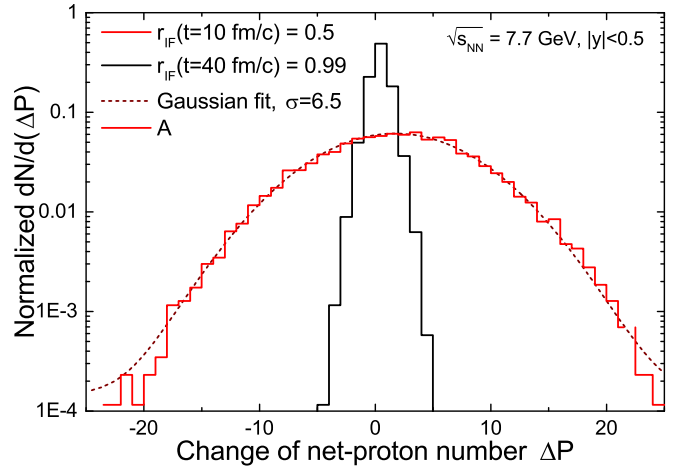}
\caption{[Color online] Probability distribution of the change of the net-proton number in the time interval between $t=10$ fm/c (red line), $t=40$ fm/c (black line) and the last scattering. We show results for $\sqrt{s_{\rm NN}}= 7.7$ GeV. The corresponding correlation values $r_{\rm IF}(t)$ are 0.5 for $t=10$ and 0.99 for $t=40$ fm/c. We also fitted the distribution for $t=10$ fm/c with a Gaussian function and obtained a width $\sigma=6.5$ which is close to the width of the final net-proton distribution $\sigma_{\rm final} = 7.5$. 
}\label{f5}
\end{figure}		%       ----------------------------------------- 

\begin{table}[t]
\centering
\begin{tabular}{|c|c|}
\hline
Beam energy $\sqrt{s_{\mathrm{NN}}}$ [GeV] & Gaussian width \\
\hline \hline
7.7 &  6.5 \\ \hline
11 &  6.3 \\ \hline
200 &  6.9 \\ \hline
\end{tabular}
\caption{Gaussian widths of the change of net-proton number, during the full hadronic rescattering, for different beam energies}
\label{t1}
\end{table} 

As a last step we will demonstrate the effect of the de-correlation of the different net-charge and net-particle numbers on their final number distributions. 
To do so we look at the distribution of the change of a particular net-particle number (i.e. the net-proton number) for different values of $r_{\rm IF}(t)$ as extracted from our simulations. The change of the net-proton number $\Delta P$ is defined as the final net-proton number in a single event minus the net-proton number at a given time $t$ in that single event. If at time $t$ the event has already its final net-proton number we find, $r_{\rm IF}(t)=1$, and the distribution of $\Delta P$ would be a delta function with a possible shift along the $\Delta P$-axis. The smaller the value of $r_{\rm IF}(t)$ becomes, the broader we expect the distribution of the change to be. Figure \ref{f5} shows this distribution function for two different values of $r_{\rm IF}(t)$, at times $t=10$ fm/c and $t=40$ fm/c, for central collisions at a beam energy of $\sqrt{s_{\rm NN}}= 7.7$ GeV. We observe that already for a small de-correlation $r_{\rm IF}(t=40)=0.99$, we obtain a distribution which does not closely resemble a delta function but rather a Gaussian with finite width, even though the width is very small compared to the mean net-proton number which is 46.
However, for a correlation of $r_{\rm IF}(t=10)=0.5$, the change of the net-proton number has a broad distribution. We were able to fit this distribution with a Gaussian of width $\sigma=6.5$, which is only slightly smaller than the width of the final net-proton number distribution which is 7.5. 
Finally, table \ref{t1} shows the width of the gaussian fit to the change of the net proton number for 
different beam energies. The width of the Gaussian smearing is approximately constant for all beam energies, 
which is consistent with the observation in figure \ref{f3} that the de-correlation is almost independent of 
beam energy.

\section{Relevance for observations}
At this point we want to put our results in context with previous publications \cite{Braun-Munzinger:2014lba,Alba:2015iva} on interpreting data from heavy ion collisions. It is important to keep in mind that, even though we have shown that the actual net-charge number in an acceptance window will change in every event due to the rescattering, this does not necessarily mean that also the measured cumulant of that net-charge will also change. Let us take for example a system which does not entail any correlations between particles. Then the cumulants of any order are merely the mean number of that charge (see e.g. \cite{Bzdak:2016sxg}). So if the mean particle number is fixed at some point during the evolution the cumulants for a non correlated system will be fixed by definition at that same point, called the chemical freeze out point (not to be mistaken with the latest point of chemical equilibrium).

If the system has strong correlations present at the point where the mean particle numbers are fixed, due to a phase transition or critical endpoint, and these correlations are washed out by the hadronic rescattering, again the cumulants will simply reflect the mean values. Note that there is a small contribution to the cumulants from conservation laws. These anti-correlation effects are small for high beam energies and they essentially cancel when taking ratios of same order cumulants, thus they are not relevant for the discussion presented here.

In consequence, if indeed the determination of the 'freeze-out' point by use of net-charge cumulants is consistent with the 'freeze-out' point of the mean particle values, one can only conclude that either no correlations where present near the 'freeze-out' or they have been washed out by the rescattering. If the measurement shows practically independent particle production then no new information can be gained from the measurement of higher order cumulants as compared to the mean particle multiplicities.

\section{Summary}

We have explored the time dependence of the correlation function of conserved charges in Au+Au reactions at various beam energies. We found that the hadronic rescattering phase leads to a substantial decorrelation of the conserved charge distributions. 
This effect is present for all investigated conserved charges and is mostly energy independent. 
This means that the final distribution function, the only one which can be observed, is to a large extent uncorrelated to that of the newly born hadrons which may carry information on the quark phase of QCD. Hence, a naive comparison of the experimentally measured correlations with the values as calculated by lattice QCD, appropriate for a stationary and infinite model system only, can produce unjustified and misleading conclusions on the phase structure of QCD.

\section{Acknowledgments}
V.V. appreciates the support from HGS-HIRe for FAIR.
H.St. appreciates the support from J.M.~Eisenberg Laureatus chair.
The computational resources were provided by the LOEWE Frankfurt Center for Scientific Computing (LOEWE-CSC), 
and by the Kronos computing cluster at GSI.

%%%%%%%%%%%%%%%%%%%%%%%%%%%%%%%%%%%%%%%%%%%%%%%%%%%%%%%%%%%%%%%%%%%%%%%%%%%%%%%
%%%%%%%%%%%%%%%%%%%%%%%%%%%%%%%%%%%%%%%%%%%%%%%%%%%%%%%%%%%%%%%%%%%%%%%%%%%%%%% 

\end{document}